\documentclass[reprint, aps, amsmath, amssymb, twocolumn, superscriptaddress, longbibliography]{revtex4-2}

\usepackage[utf8]{inputenc}
\usepackage{graphicx}
\usepackage{xcolor} 
\usepackage{amsmath} 
\DeclareMathOperator{\Tr}{Tr}
\usepackage{textgreek} 
\usepackage{braket} 
\usepackage[separate-uncertainty=true]{siunitx} 
\DeclareSIUnit\gauss{G}

\begin{document}
\title{Protecting qubit coherence by spectrally engineered driving of the spin environment}
\author{Maxime Joos}
\affiliation{Department of Physics, University of California, Santa Barbara, California 93106, USA}
\author{Dolev Bluvstein}
\affiliation{Department of Physics, University of California, Santa Barbara, California 93106, USA}
\affiliation{Department of Physics, Harvard University, Cambridge, MA 02138, USA}
\author{Yuanqi Lyu}
\affiliation{Department of Physics, University of California, Santa Barbara, California 93106, USA}
\affiliation{Department of Physics, University of California, Berkeley, CA 94720, USA}
\author{David Weld}
\affiliation{Department of Physics, University of California, Santa Barbara, California 93106, USA}
\author{Ania Bleszynski Jayich}
\affiliation{Department of Physics, University of California, Santa Barbara, California 93106, USA}

\begin{abstract}
Modern quantum technologies rely crucially on techniques to mitigate quantum decoherence; these techniques can be either passive, achieved for example via materials engineering, or active, typically achieved via pulsed monochromatic driving fields applied to the qubit.
Using a solid-state defect spin coupled to a microwave-driven spin bath, we experimentally demonstrate a decoherence mitigation method based on spectral engineering of the environmental noise with a polychromatic drive waveform, and show that it outperforms monochromatic techniques.
Results are in agreement with quantitative modeling, and open the path to active decoherence protection using custom-designed waveforms applied to the environment rather than the qubit.
\end{abstract}

\maketitle

Quantum decoherence, resulting from the unavoidable coupling between a qubit and its environment, underlies fundamental descriptions of the quantum-classical transition \cite{Zurek2007, Schlosshauer2019} and poses a major challenge to a variety of quantum technologies \cite{Acin2018}.
Several methods exist to mitigate environment-induced decoherence.
Materials-based approaches aim to perfect the environment through techniques like surface passivation and optimized synthesis~\cite{Quintana2014, Kumar2016, ohno2012, Eichhorn2019}.
Dynamical decoupling instead relies on manipulating the qubit rapidly enough to average out deleterious environmental fluctuations at particular frequencies~\cite{Viola1998, Viola1999, Bylander2011, deLange2010}.
However, the flexibility and power of this technique comes at some cost to the qubit's utility, as the decoupling pulses need to be interleaved with gate operations or sensing sequences~\cite{Vandersar2012}.
An alternative and complementary approach aims at manipulating the noise frequency spectrum of the bath itself, a technique referred to as spin-bath driving~\cite{Barry2020} and originally developed in nuclear magnetic resonance and known as spin decoupling~\cite{Slichter1990}.
Recent work has shown that monochromatic driving of a spin bath can extend the coherence of bulk~\cite{Bauch2018} and near-surface NV centers~\cite{Dolev2019, ishizu2020spin}.
An intrinsic limitation of this monochromatic approach is that it does not efficiently address spectrally broad classes of spins that naturally emerge in inhomogeneous environments and interacting spin baths~\cite{Ernst1966}.

In this work, we introduce a new polychromatic drive scheme and demonstrate experimentally that spectrally engineered driving of a spin bath enhances qubit coherence beyond the limit of what can be achieved with monochromatic driving.
The drive operates in analogy to motional narrowing and broad-band decoupling \cite{Ernst1966} in nuclear magnetic resonance: driving the spin bath accelerates incoherent bath fluctuations, reducing the integrated phase acquired from the spin bath.
Our driving scheme not only enables significantly increased power efficiency for protecting coherence, it also paves the way toward more complex and powerful techniques of tailored dynamical bath engineering.

\begin{figure}[!tb]
    \center{\includegraphics[width=\linewidth]
    {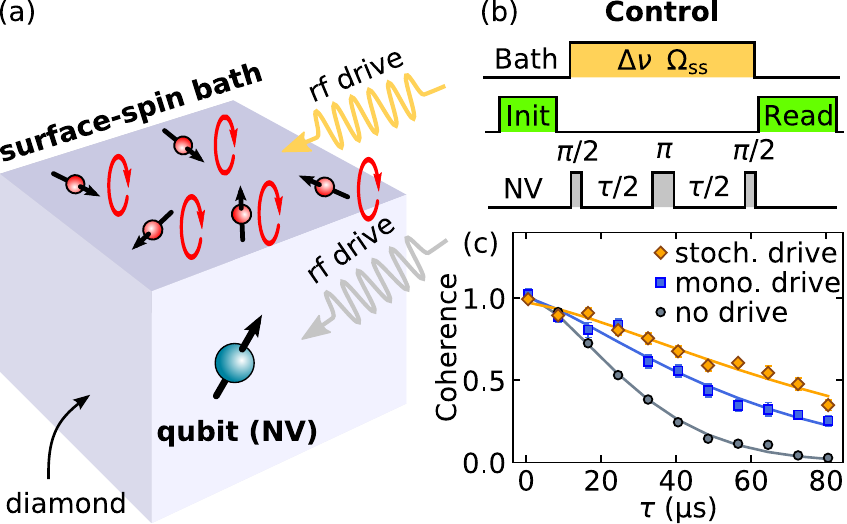}}
    \caption{\textbf{Coherence extension by driving the spin bath}.
    \textbf{(a)} A shallow NV center (qubit) is dephased by the magnetic noise originating from a bath of surface spins.
    \textbf{(b)} Hahn-echo sequence used to measure coherence during bath driving with linewidth $\Delta \nu$ and Rabi frequency $\Omega_{ss}$.
    \textbf{(c)} Coherence decay without driving (grey), with monochromatic driving (blue) and polychromatic/stochastic driving (orange), described in the text.
    Solid lines are fits to $\exp{\left[-(\tau/T_2)^n\right]}$.
    }
    \label{fig:1}
\end{figure}

Fig. \ref{fig:1}(a) shows a model of the system we investigate. The polychromatic drive is applied to the spin environment of a shallow nitrogen-vacancy (NV) center in diamond (see \cite{supplemental} for details on the diamond sample preparation).
The NV center is a solid-state qubit that exhibits long coherence in ambient conditions and is being used in a variety of applications ranging from networking to quantum sensing.
This work uses NV centers located just a few nanometers below the diamond surface.
Recent studies have identified magnetic noise from the surface spins as a major contributor to shallow NV decoherence~\cite{Mamin2012, Grinolds2014, Dolev2019, Myers2014, Sangtawesin2019, Stacey2019}.
Radio-frequency (rf) fields, delivered by a single free-space antenna, enable coherent control of NV qubit states $\{\ket{m_s=0}, \ket{m_s = -1}\}$ and surface spin qubit states $\{\ket{\uparrow}, \ket{\downarrow}\}$. NV centers are addressed individually by means of a home-built confocal microscope operating in ambient conditions, and are subjected to a static magnetic field $B_0 \approx \text{\SI{315}{\gauss}}$ aligned along the NV axis.
We use green ($532\text{-nm}$ wavelength) laser pulses for initialization and readout of the NV spin state.
The NV coherence $C(\tau)$ is probed using the Hahn-echo sequence represented in Fig. \ref{fig:1}(b); simultaneously, the surface spins are driven with spectrally engineered rf fields.
Specifically, we apply drives with Lorentzian spectral line-shapes characterized by a tunable full-width at half maximum (FWHM) $\Delta \nu$, generated by stochastic phase modulation of a carrier wave with Rabi frequency $\Omega_{ss}$ on the surface spin transition, denoted as a \textit{stochastic drive} (see \cite{supplemental} for details).
Spectrum engineering via phase modulation (as opposed to amplitude modulation) is a key feature of the experiment, ensuring that the driving power remains constant, whereas amplitude modulation could cause fluctuating AC Stark shifts of the NV spin states and decohere the NV \cite{Dolev2019}.
The NV coherence time $T_2$ is extracted from a stretched exponential fit  ($\exp{\left[-(\tau/T_2)^n\right]}$) to the measured coherence decay shown in Fig. \ref{fig:1}(c).

\begin{figure}[tbp]
\center{\includegraphics[width=\linewidth]{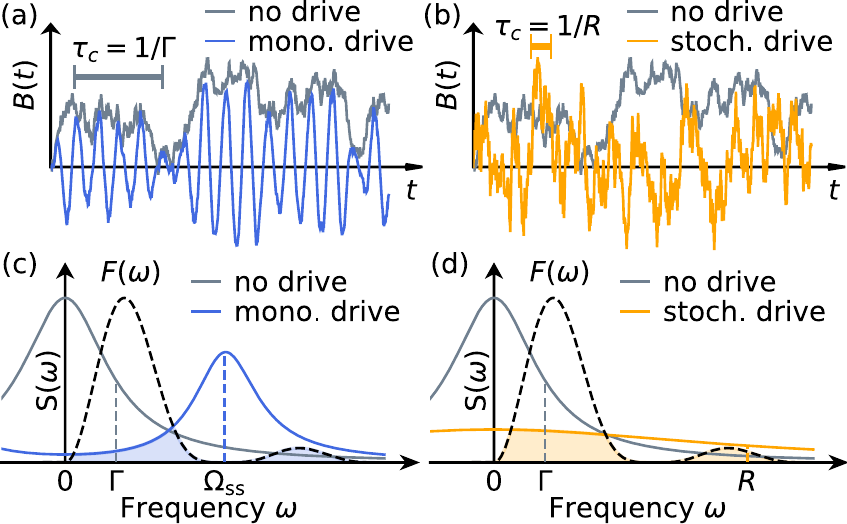}}
\caption{\textbf{Principle of monochromatic and stochastic bath driving}.
Time domain representations of the magnetic noise $B(t)$ produced by a spin bath, (a) for monochromatic driving (blue) and (b) for stochastic driving (orange).
For simplicity, all the spins are assumed to have the same resonance frequency.
Frequency domain representations of the noise spectral density $S(\omega)$, (c) for monochromatic driving (blue) and (d) for stochastic driving (orange).
Both driving methods reduce the overlap between the filter function $F(\tau, \omega)$ (dashed line) and the noise spectrum, as compared to the undriven case (grey curve), resulting in increased coherence.
}
\label{fig:2}
\end{figure}

To understand the effect of bath driving on the qubit's coherence, we first give a quantitative description of NV decoherence based on a model of the surrounding bath.
The NV coherence $C(\tau)$ is determined by the overlap between a filter function and the noise spectral density \cite{deSousa2009, Biercuk2011, Degen2017, Cywinski2008}:
\begin{equation}
    C(\tau) = \exp\left[-\frac{1}{\pi}\int_0^{\infty} d\omega S(\omega) F(\tau, \omega)\right],
\label{eq:1}
\end{equation}
where
\begin{equation}
    S(\omega) = \gamma^2 \int_{-\infty}^{\infty} e^{-i\omega t} \langle B(t'+t) B(t') \rangle dt,
\label{eq:spectrum1}
\end{equation}
is the magnetic noise spectral density experienced by the NV (see Fig. \ref{fig:2}(c) and (d)), $\gamma$ is the gyromagnetic ratio, $B$ is the field component along the NV axis, $F(\tau, \omega) = 8/\omega^2 \sin^4{(\omega \tau/4)}$ is the qubit Hahn-echo filter function, and $\omega$ is the angular frequency. This decoherence model assumes a random Gaussian distribution for the noise amplitudes and is valid in the regime of pure dephasing.

Our strategy to decouple the qubit from its noise environment relies on actively reshaping the spectral density so that it minimally overlaps with the filter function.
This decoupling mechanism and its underlying microscopic model is illustrated in Fig. \ref{fig:2}.
The upper panels are time-domain representations of the magnetic noise $B(t)$ produced by the surface spins, and the lower panels are the corresponding frequency domain representations $S(\omega)$.
We now illustrate three cases of driving: no driving, monochromatic, and stochastic. 

In the absence of driving (grey curve in Fig. \ref{fig:2}), intrinsic spin relaxations at a rate $\Gamma$ lead to a random process for the noise, characterized by a correlation time $\tau_c = 1/\Gamma$ in the time domain (Fig. \ref{fig:2}(a) and (b)) and a Lorentzian spectrum with half width at half maximum (HWHM) $\Gamma$ centered at $\omega=0$ (Fig. \ref{fig:2}(c) and (d)).
The overlap of the noise spectrum with the Hahn-echo filter function sets the contribution of surface spins to NV decoherence.

In the case of resonant monochromatic bath driving with a Rabi frequency $\Omega_{ss}$, the surface spins undergo Rabi flopping in addition to their intrinsic relaxation dynamics, leading to oscillations of the field $B(t)$ in the time domain (blue curve in Fig. \ref{fig:2}(a)).
In the frequency domain (Fig. \ref{fig:2}(c)), monochromatic driving shifts the center of the spectral density from 0 to $\Omega_{ss}$;
the overlap with the qubit filter function (represented by the blue shaded area) is consequently reduced from the undriven case.

In contrast, stochastic driving consists of a continuum of frequency tones with random phase relations.
Each tone drives partial Rabi oscillations of the surface spins and the incoherent sum of these Rabi oscillations results in random evolution as shown in Fig. \ref{fig:2}(b).
These incoherent dynamics lead to an effective relaxation rate $R$ induced by stochastic driving, on top of the intrinsic relaxation rate $\Gamma$ \cite{supplemental}.
In the case of a broad Lorentzian driving spectrum $\Delta \nu \gg \Omega_{ss}, \Gamma$,
\begin{equation}
    R = \Gamma + 2\frac{\Omega_{ss}^2}{\Delta \nu}.
    \label{eq:2}
\end{equation}
In the frequency domain (Fig. \ref{fig:2}(d)), the effect of the stochastic drive is to broaden and flatten the noise spectral density.
The overlap with the filter function is reduced, leading to an extension of coherence.
Compared to monochromatic driving, the advantage of stochastic driving is that it more efficiently addresses a broad spectral range of spins: the spectral range over which spins are decoupled scales linearly (quadratically) with the Rabi frequency $\Omega_{ss}$ for monochromatic (stochastic) driving \cite{Ernst1966}.

\begin{figure}[!tb]
    \center{\includegraphics[width=\linewidth]
    {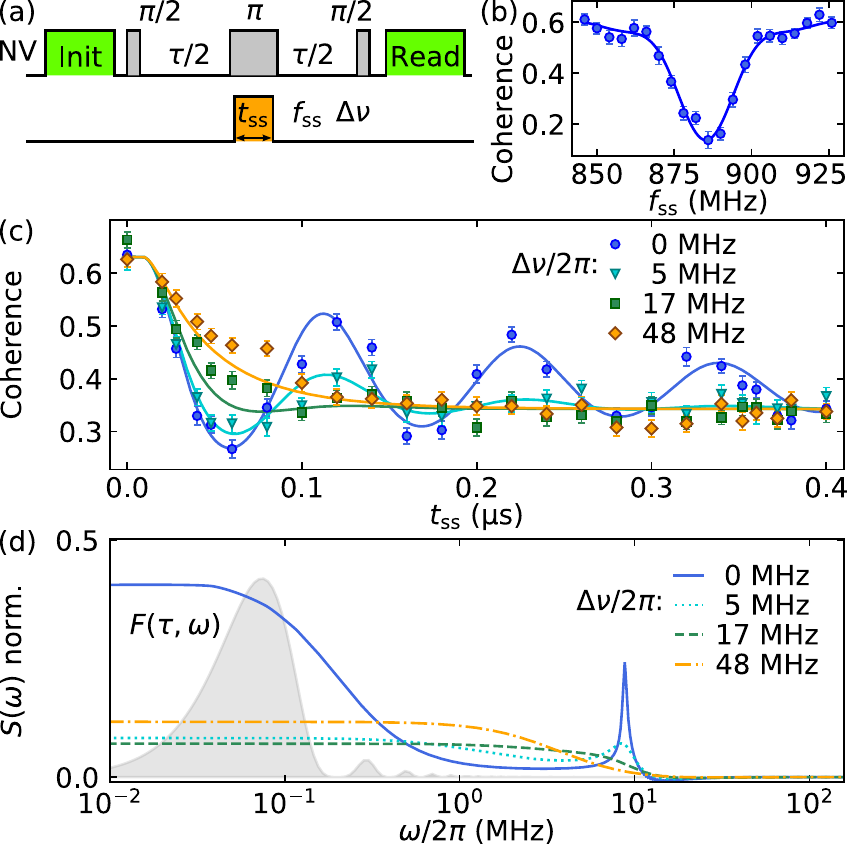}}
    \caption{\textbf{Probing bath correlations to quantify noise spectrum under drive.}
    (a) Pulse sequence used to probe the surface spin bath, and 
    (b) corresponding spectrum with $\Delta \nu = 0$, $t_{ss} = 60 \text{ ns}$, $\tau = \text{\SI{20}{\micro\second}}$, $\Omega_{ss}/2\pi = 9.7\text{ MHz}$, and an NV $\pi\text{-pulse}$ of \SI{80}{\nano\second}. 
    (c) Experimental spin bath dynamics, obtained by measuring NV coherence as a function of pulse length $t_{ss}$ for different excitation linewidths $\Delta\nu$ and for $\Omega_{ss}/2\pi = 8.7 \text{ MHz}$.
    Increasing $\Delta \nu$ accelerates the damping of the Rabi oscillations, eventually reaching an overdamped regime.
    (d) Noise spectral densities obtained by Fourier transformation of the fitted DEER signals in (c) and normalized so that $2\int_{0}^{\infty} S(\omega) d\omega/2\pi = 1$. Grey shaded area represents the Hahn-echo filter function corresponding to a sequence duration of $\tau =\text{\SI{10}{\micro\second}}$.
    }
    \label{fig:3}
\end{figure}

Having discussed the predicted spin dynamics under stochastic excitation and its potential for extending qubit coherence, we now experimentally characterize the frequency and time response of the surface-spin bath and the effects of stochastic driving.
We use the NV to probe the surface spins dynamics with the double electron-electron resonance (DEER) sequence shown in Fig. \ref{fig:3}(a).
The Hahn-echo sequence on the NV cancels out low frequency noise from the environment, but the DEER pulse (duration $t_{ss}$, center frequency $f_{ss}$ and spectral width $\Delta\nu$) selectively recouples the surface spins and reveals their dynamics.

Figure \ref{fig:3}(b) shows the NV coherence as a function of $f_{ss}$, where the fixed pulse duration $t_{ss} = 60\text{ ns}$ is chosen to be a $\pi\text{-pulse}$ on resonance with the surface spins.
The acquired spectrum exhibits a resonance at \SI{885.3 +- 0.4}{\mega\hertz} which is the magnetic signature of a $g=2$ $\text{spin-}1/2$ particle in the applied static field of \SI{315}{\gauss}.
The line-shape is best modeled by assuming inhomogeneous broadening of the surface spins and accounting for the finite pulse width $t_{ss}$ (see \cite{supplemental} for details).
From the fit, we extract the FWHM of the underlying Gaussian distribution of the spins $2\Gamma_2 = \text{\SI{15.7 +- 1.3}{\mega\hertz}}$.

Having confirmed the electronic nature of the surface spins and identified their inhomogeneous broadening, we now investigate the temporal response of the bath.
Figure \ref{fig:3}(c) shows the measured NV coherence as a function of $t_{ss}$ for different line-widths $\Delta \nu$ of the drive and a fix sequence duration $\tau = \text{\SI{16}{\micro\second}}$.
When $\Delta \nu =0$, the NV coherence exhibits damped oscillations as a function of $t_{ss}$.
The oscillations reflect the Rabi flopping of the surface spins while the damping is due to the inhomogeneous broadening and the intrinsic relaxation.
For $\Delta\nu > 0$, the stochastic DEER pulse decorrelates the field produced by the surface spins between the first and second half of DEER sequence, leading to increased damping of the Rabi oscillations.
For $\Delta\nu/2\pi = 48\text{ MHz} > 2\Gamma_{2}$, the measured NV coherence decays exponentially at a rate given by Eq. (\ref{eq:2}), as expected for broad Lorentzian excitation.
The solid curves in Fig. \ref{fig:3}(c) are fits using our model of inhomogeneously broadened surface spins subjected to a stochastic drive \cite{supplemental}.
From these fits, we extract an independent measure of the inhomogeneous broadening $2\Gamma_2 = \text{\SI{14.9 +- 1.2}{\mega\hertz}}$ in agreement with the DEER spectrum in Fig. \ref{fig:3}(b), and a Rabi frequency $\Omega_{ss}/2\pi = \text{\SI{8.7 +- 0.1}{\mega\hertz}}$.

To investigate the noise spectrum generated by the measured surface spin dynamics, Figure \ref{fig:3}(d) shows the Fourier transform of the fitted bath correlations of Fig. \ref{fig:3}(c) obtained after symmetrization with respect to $t_{ss} = 0$.
We constrain the integrated power of the spectra to be constant.
As expected from the time domain dynamics, the spectrum under monochromatic driving ($\Delta \nu = 0$) exhibits a peak at $\Omega_{ss}/2\pi$, but still has a remaining static component due to the detuned Rabi oscillations of the inhomogeneously broadened spins (the simplified Fig. \ref{fig:2} schematic does not account for this inhomogeneous broadening).
In the case of stochastic driving, the noise spectrum is flattened and broadened, and the overlap with $F(\tau,\omega)$ is considerably reduced.

\begin{figure}[!tb]
    \center{\includegraphics[width=85.7mm]
    {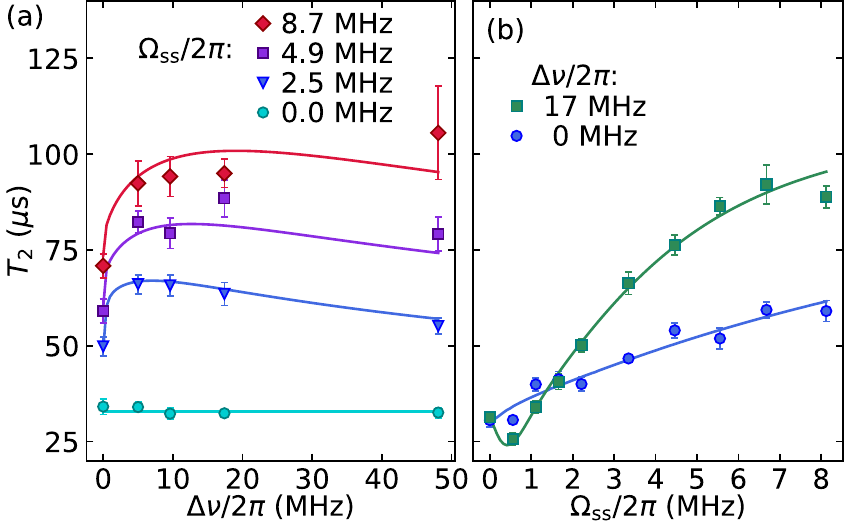}}
    \caption{\textbf{Coherence time for different driving parameters}.
    (a) NV $T_2$ as a function of $\Delta \nu$ for different bath Rabi frequencies $\Omega_{ss}$.
    Stochastic drive outperforms monochromatic drive ($\Delta \nu =0$).
    (b) $T_2$ as a function of drive Rabi frequency $\Omega_{ss}$ for monochromatic driving $\Delta \nu = 0$ and stochastic driving with $\Delta \nu = 17 \text{ MHz}$.
    Solid lines are fit using our model \cite{supplemental}.
    }
    \label{fig:4}
\end{figure}

We now demonstrate that stochastic driving extends NV coherence beyond the gains of monochromatic bath driving.
We perform the sequence shown in Fig. \ref{fig:1}(b) and measure the NV coherence time $T_2\left(\Omega_{ss}, \Delta \nu \right)$. We observe that stochastic driving outperforms monochromatic driving over a broad range of drive parameters.
Figure \ref{fig:4} shows the coherence time $T_2(\Omega_{ss}, \Delta\nu)$, where the bare coherence time $T_{2}(\Omega_{ss}=0) = \text{\SI{33.1 +- 1.5}{\micro\second}}$.
Figures \ref{fig:4}(a) and (b) highlight the $\Delta\nu$ and $\Omega_{ss}$ dependences, respectively.
For example, we measure a 2.7-fold increase in coherence time with stochastic driving at $(\Omega_{ss}/2\pi = 4.9\text{ MHz}, \Delta\nu/2\pi = 17.4\text{ MHz})$, as compared to a 1.8-fold increase with monochromatic driving at the same power.
Our experimental data are well-captured by Eq.~(\ref{eq:1}) where $S(\omega)$ includes contributions from two g=2 spin baths (both are driven, but have different bath parameters), and a residual bath which is undriven (see \cite{supplemental} for details).

Our results suggest that, for a given driving power, an optimal $\Delta\nu$ exists that maximizes the coherence. The non-monotonic dependence on $\Delta\nu$ can be understood intuitively as follows:
starting from $\Delta\nu = 0$, increasing $\Delta\nu$ enables a larger fraction of the inhomogeneously broadened surface spins to be resonantly addressed and therefore decoupled from the NV center.
When $\Delta\nu$ exceeds the inhomogeneous broadening of the surface spins (reported above as $2\Gamma_2 \sim  16 \text{ MHz}$), drive energy is spread over non-resonant frequencies and the power efficiency of stochastic driving decreases.

Figure \ref{fig:4}(b) also shows that stochastic driving reduces coherence at the lowest Rabi frequencies, an effect also captured by our model.
Intuitively, as $\Omega_{ss}$ first increases from zero, the noise spectral density broadens, increasing in overlap with the filter function (see Fig. \ref{fig:2}(d)) and decreasing the coherence, before broadening well beyond the peak and flattening out.
In \cite{supplemental} we derive the condition for stochastic driving to increase coherence: $\sqrt{\Gamma R} \gtrsim 2\pi/T_2$.
Incorporating two driven baths with long- and short-lived correlations (as suggested by \cite{Tetienne2018}) in our model is essential to quantitatively capture the features of Fig. \ref{fig:4}(a) and (b), including the drop of coherence discussed above as detailed in the supplemental \cite{supplemental}.
From the fit in Fig. \ref{fig:4}(b), we extract the correlation times of the two driven baths: \SI{68 \pm 17}{\micro\second} and \SI{0.37 +- 0.05}{\micro\second}.

Finally, we provide a qualitative explanation for the gains of stochastic driving relative to monochromatic driving.
In order to completely decouple an ensemble of inhomogeneously broadened spins, monochromatic driving relies on power broadening and requires $\Omega_{ss} \gg \Gamma_2$.
On the other hand, stochastic driving naturally addresses all the spins as long as $\Delta\nu \gtrsim \Gamma_2$.
Combining this relation with Eq. (\ref{eq:2}), the condition for decoupling all the spins in the case of stochastic driving is $\Omega_{ss} \gg \sqrt{\Gamma \Gamma_2}$.
Therefore, stochastic driving outperforms monochromatic driving as long as $\Gamma < \Gamma_2$, which is often satisfied and, for example, is satisfied by several orders of magnitude for near-surface NV qubits.
For the surface spin bath investigated here, we observe $\Gamma_2 \sim \text{\SI{10}{\mega\hertz}}$, and expect $\Gamma \sim \SI{0.03}{\mega\hertz}$ as reported in \cite{Sushkov2014} (consistent with our observations), explaining the better performance of stochastic driving as $\Gamma \ll \Gamma_2$.

A practical advantage of stochastic driving is that it reduces the driving power required to achieve a given coherence extension.
AC Stark shifts of the qubit energy levels and heating at high rf driving powers are experimental limitations to the coherence time that scale with drive power ($\propto |\Omega_{ss}|^2$) \cite{Dolev2019}. 
For example, Fig. \ref{fig:4}(b) shows that doubling the coherence necessitates $\Omega_{ss}/2\pi \approx \text{\SI{3}{\mega\hertz}}$ for stochastic driving, and $\approx \text{\SI{8}{\mega\hertz}}$ for monochromatic driving, corresponding to a 7-fold power reduction when stochastic driving is used.
This advantage may prove crucial for avoiding heating and other undesired effects, in particular for cryogenic systems like superconducting qubits \cite{Chang2013}, which could also benefit from spin bath driving techniques.
While coherent driving of surface electric dipoles that decohere trapped ion qubits \cite{Brownnutt2015, Daniilidis2011, Hite2013, Noel2019} or NV qubits \cite{Myers2017} may be infeasible, this work  suggests that such dipoles could possibly be incoherently driven and motionally narrowed.

In this work we have considered only driving with a Lorentzian spectrum, but other types of polychromatic driving could be even more advantageous.
The drive could for example be engineered to precisely cover the exact spectral shape of the noise sources, and engineering alternative spectral shapes could also enable maximization of the coherence at a desired Rabi frequency.
In the supplemental material \cite{supplemental}, we show that a Gaussian spectrum for the drive performs better than a Lorentzian spectrum at specific Rabi frequencies.
Ultimately, our method may be extended to polychromatic driving with non-random phase relations between different frequencies.

In conclusion, we demonstrate the extension of coherence of a qubit by driving its spin environment with spectrally engineered fields.
We show that stochastic driving of the surface-spin bath of shallow NV centers is more efficient than monochromatic driving, and corroborate our findings with a quantitative model.
Further improvements, especially for mesoscopic baths, may be possible using customized structured drive waveforms.

We thank Shreyas Parthasarathy and Esat Kondakci for careful reading of the manuscript.
We gratefully acknowledge support from the US Department of Energy (BES grant No. DE-SC0019241) for surface spin studies and the DARPA DRINQS program (Agreement No. D18AC00014) for driving protocols.
D.B. acknowledges support from the NSF Graduate Research Fellowship Program (grant DGE1745303) and The Fannie and John Hertz Foundation.

\bibliography{bibliography}

\begin{thebibliography}{36}%
\makeatletter
\providecommand \@ifxundefined [1]{%
 \@ifx{#1\undefined}
}%
\providecommand \@ifnum [1]{%
 \ifnum #1\expandafter \@firstoftwo
 \else \expandafter \@secondoftwo
 \fi
}%
\providecommand \@ifx [1]{%
 \ifx #1\expandafter \@firstoftwo
 \else \expandafter \@secondoftwo
 \fi
}%
\providecommand \natexlab [1]{#1}%
\providecommand \enquote  [1]{``#1''}%
\providecommand \bibnamefont  [1]{#1}%
\providecommand \bibfnamefont [1]{#1}%
\providecommand \citenamefont [1]{#1}%
\providecommand \href@noop [0]{\@secondoftwo}%
\providecommand \href [0]{\begingroup \@sanitize@url \@href}%
\providecommand \@href[1]{\@@startlink{#1}\@@href}%
\providecommand \@@href[1]{\endgroup#1\@@endlink}%
\providecommand \@sanitize@url [0]{\catcode `\\12\catcode `\$12\catcode
  `\&12\catcode `\#12\catcode `\^12\catcode `\_12\catcode `\%12\relax}%
\providecommand \@@startlink[1]{}%
\providecommand \@@endlink[0]{}%
\providecommand \url  [0]{\begingroup\@sanitize@url \@url }%
\providecommand \@url [1]{\endgroup\@href {#1}{\urlprefix }}%
\providecommand \urlprefix  [0]{URL }%
\providecommand \Eprint [0]{\href }%
\providecommand \doibase [0]{http://dx.doi.org/}%
\providecommand \selectlanguage [0]{\@gobble}%
\providecommand \bibinfo  [0]{\@secondoftwo}%
\providecommand \bibfield  [0]{\@secondoftwo}%
\providecommand \translation [1]{[#1]}%
\providecommand \BibitemOpen [0]{}%
\providecommand \bibitemStop [0]{}%
\providecommand \bibitemNoStop [0]{.\EOS\space}%
\providecommand \EOS [0]{\spacefactor3000\relax}%
\providecommand \BibitemShut  [1]{\csname bibitem#1\endcsname}%
\let\auto@bib@innerbib\@empty
\bibitem [{\citenamefont {Zurek}(2007)}]{Zurek2007}%
  \BibitemOpen
  \bibfield  {author} {\bibinfo {author} {\bibfnamefont {W.~H.}\ \bibnamefont
  {Zurek}},\ }\enquote {\bibinfo {title} {Decoherence and the transition from
  quantum to classical --- revisited},}\ in\ \href {\doibase
  10.1007/978-3-7643-7808-0_1} {\emph {\bibinfo {booktitle} {Quantum
  Decoherence: Poincar{\'e} Seminar 2005}}},\ \bibinfo {editor} {edited by\
  \bibinfo {editor} {\bibfnamefont {B.}~\bibnamefont {Duplantier}}, \bibinfo
  {editor} {\bibfnamefont {J.-M.}\ \bibnamefont {Raimond}}, \ and\ \bibinfo
  {editor} {\bibfnamefont {V.}~\bibnamefont {Rivasseau}}}\ (\bibinfo
  {publisher} {Birkh{\"a}user Basel},\ \bibinfo {address} {Basel},\ \bibinfo
  {year} {2007})\ pp.\ \bibinfo {pages} {1--31}\BibitemShut {NoStop}%
\bibitem [{\citenamefont {Schlosshauer}(2019)}]{Schlosshauer2019}%
  \BibitemOpen
  \bibfield  {author} {\bibinfo {author} {\bibfnamefont {M.}~\bibnamefont
  {Schlosshauer}},\ }\href {\doibase
  https://doi.org/10.1016/j.physrep.2019.10.001} {\bibfield  {journal}
  {\bibinfo  {journal} {Physics Reports}\ }\textbf {\bibinfo {volume} {831}},\
  \bibinfo {pages} {1 } (\bibinfo {year} {2019})},\ \bibinfo {note} {quantum
  decoherence}\BibitemShut {NoStop}%
\bibitem [{\citenamefont {Ac{\'{\i}}n}\ \emph {et~al.}(2018)\citenamefont
  {Ac{\'{\i}}n}, \citenamefont {Bloch}, \citenamefont {Buhrman}, \citenamefont
  {Calarco}, \citenamefont {Eichler}, \citenamefont {Eisert}, \citenamefont
  {Esteve}, \citenamefont {Gisin}, \citenamefont {Glaser}, \citenamefont
  {Jelezko}, \citenamefont {Kuhr}, \citenamefont {Lewenstein}, \citenamefont
  {Riedel}, \citenamefont {Schmidt}, \citenamefont {Thew}, \citenamefont
  {Wallraff}, \citenamefont {Walmsley},\ and\ \citenamefont
  {Wilhelm}}]{Acin2018}%
  \BibitemOpen
  \bibfield  {author} {\bibinfo {author} {\bibfnamefont {A.}~\bibnamefont
  {Ac{\'{\i}}n}}, \bibinfo {author} {\bibfnamefont {I.}~\bibnamefont {Bloch}},
  \bibinfo {author} {\bibfnamefont {H.}~\bibnamefont {Buhrman}}, \bibinfo
  {author} {\bibfnamefont {T.}~\bibnamefont {Calarco}}, \bibinfo {author}
  {\bibfnamefont {C.}~\bibnamefont {Eichler}}, \bibinfo {author} {\bibfnamefont
  {J.}~\bibnamefont {Eisert}}, \bibinfo {author} {\bibfnamefont
  {D.}~\bibnamefont {Esteve}}, \bibinfo {author} {\bibfnamefont
  {N.}~\bibnamefont {Gisin}}, \bibinfo {author} {\bibfnamefont {S.~J.}\
  \bibnamefont {Glaser}}, \bibinfo {author} {\bibfnamefont {F.}~\bibnamefont
  {Jelezko}}, \bibinfo {author} {\bibfnamefont {S.}~\bibnamefont {Kuhr}},
  \bibinfo {author} {\bibfnamefont {M.}~\bibnamefont {Lewenstein}}, \bibinfo
  {author} {\bibfnamefont {M.~F.}\ \bibnamefont {Riedel}}, \bibinfo {author}
  {\bibfnamefont {P.~O.}\ \bibnamefont {Schmidt}}, \bibinfo {author}
  {\bibfnamefont {R.}~\bibnamefont {Thew}}, \bibinfo {author} {\bibfnamefont
  {A.}~\bibnamefont {Wallraff}}, \bibinfo {author} {\bibfnamefont
  {I.}~\bibnamefont {Walmsley}}, \ and\ \bibinfo {author} {\bibfnamefont
  {F.~K.}\ \bibnamefont {Wilhelm}},\ }\href {\doibase 10.1088/1367-2630/aad1ea}
  {\bibfield  {journal} {\bibinfo  {journal} {New Journal of Physics}\ }\textbf
  {\bibinfo {volume} {20}},\ \bibinfo {pages} {080201} (\bibinfo {year}
  {2018})}\BibitemShut {NoStop}%
\bibitem [{\citenamefont {Quintana}\ \emph {et~al.}(2014)\citenamefont
  {Quintana}, \citenamefont {Megrant}, \citenamefont {Chen}, \citenamefont
  {Dunsworth}, \citenamefont {Chiaro}, \citenamefont {Barends}, \citenamefont
  {Campbell}, \citenamefont {Chen}, \citenamefont {Hoi}, \citenamefont
  {Jeffrey}, \citenamefont {Kelly}, \citenamefont {Mutus}, \citenamefont
  {O'Malley}, \citenamefont {Neill}, \citenamefont {Roushan}, \citenamefont
  {Sank}, \citenamefont {Vainsencher}, \citenamefont {Wenner}, \citenamefont
  {White}, \citenamefont {Cleland},\ and\ \citenamefont
  {Martinis}}]{Quintana2014}%
  \BibitemOpen
  \bibfield  {author} {\bibinfo {author} {\bibfnamefont {C.~M.}\ \bibnamefont
  {Quintana}}, \bibinfo {author} {\bibfnamefont {A.}~\bibnamefont {Megrant}},
  \bibinfo {author} {\bibfnamefont {Z.}~\bibnamefont {Chen}}, \bibinfo {author}
  {\bibfnamefont {A.}~\bibnamefont {Dunsworth}}, \bibinfo {author}
  {\bibfnamefont {B.}~\bibnamefont {Chiaro}}, \bibinfo {author} {\bibfnamefont
  {R.}~\bibnamefont {Barends}}, \bibinfo {author} {\bibfnamefont
  {B.}~\bibnamefont {Campbell}}, \bibinfo {author} {\bibfnamefont
  {Y.}~\bibnamefont {Chen}}, \bibinfo {author} {\bibfnamefont {I.-C.}\
  \bibnamefont {Hoi}}, \bibinfo {author} {\bibfnamefont {E.}~\bibnamefont
  {Jeffrey}}, \bibinfo {author} {\bibfnamefont {J.}~\bibnamefont {Kelly}},
  \bibinfo {author} {\bibfnamefont {J.~Y.}\ \bibnamefont {Mutus}}, \bibinfo
  {author} {\bibfnamefont {P.~J.~J.}\ \bibnamefont {O'Malley}}, \bibinfo
  {author} {\bibfnamefont {C.}~\bibnamefont {Neill}}, \bibinfo {author}
  {\bibfnamefont {P.}~\bibnamefont {Roushan}}, \bibinfo {author} {\bibfnamefont
  {D.}~\bibnamefont {Sank}}, \bibinfo {author} {\bibfnamefont {A.}~\bibnamefont
  {Vainsencher}}, \bibinfo {author} {\bibfnamefont {J.}~\bibnamefont {Wenner}},
  \bibinfo {author} {\bibfnamefont {T.~C.}\ \bibnamefont {White}}, \bibinfo
  {author} {\bibfnamefont {A.~N.}\ \bibnamefont {Cleland}}, \ and\ \bibinfo
  {author} {\bibfnamefont {J.~M.}\ \bibnamefont {Martinis}},\ }\href {\doibase
  10.1063/1.4893297} {\bibfield  {journal} {\bibinfo  {journal} {Applied
  Physics Letters}\ }\textbf {\bibinfo {volume} {105}},\ \bibinfo {pages}
  {062601} (\bibinfo {year} {2014})}\BibitemShut {NoStop}%
\bibitem [{\citenamefont {Kumar}\ \emph {et~al.}(2016)\citenamefont {Kumar},
  \citenamefont {Sendelbach}, \citenamefont {Beck}, \citenamefont {Freeland},
  \citenamefont {Wang}, \citenamefont {Wang}, \citenamefont {Yu}, \citenamefont
  {Wu}, \citenamefont {Pappas},\ and\ \citenamefont {McDermott}}]{Kumar2016}%
  \BibitemOpen
  \bibfield  {author} {\bibinfo {author} {\bibfnamefont {P.}~\bibnamefont
  {Kumar}}, \bibinfo {author} {\bibfnamefont {S.}~\bibnamefont {Sendelbach}},
  \bibinfo {author} {\bibfnamefont {M.~A.}\ \bibnamefont {Beck}}, \bibinfo
  {author} {\bibfnamefont {J.~W.}\ \bibnamefont {Freeland}}, \bibinfo {author}
  {\bibfnamefont {Z.}~\bibnamefont {Wang}}, \bibinfo {author} {\bibfnamefont
  {H.}~\bibnamefont {Wang}}, \bibinfo {author} {\bibfnamefont {C.~C.}\
  \bibnamefont {Yu}}, \bibinfo {author} {\bibfnamefont {R.~Q.}\ \bibnamefont
  {Wu}}, \bibinfo {author} {\bibfnamefont {D.~P.}\ \bibnamefont {Pappas}}, \
  and\ \bibinfo {author} {\bibfnamefont {R.}~\bibnamefont {McDermott}},\ }\href
  {\doibase 10.1103/PhysRevApplied.6.041001} {\bibfield  {journal} {\bibinfo
  {journal} {Phys. Rev. Applied}\ }\textbf {\bibinfo {volume} {6}},\ \bibinfo
  {pages} {041001} (\bibinfo {year} {2016})}\BibitemShut {NoStop}%
\bibitem [{\citenamefont {Ohno}\ \emph {et~al.}(2012)\citenamefont {Ohno},
  \citenamefont {Joseph~Heremans}, \citenamefont {Bassett}, \citenamefont
  {Myers}, \citenamefont {Toyli}, \citenamefont {Bleszynski~Jayich},
  \citenamefont {Palmstrøm},\ and\ \citenamefont {Awschalom}}]{ohno2012}%
  \BibitemOpen
  \bibfield  {author} {\bibinfo {author} {\bibfnamefont {K.}~\bibnamefont
  {Ohno}}, \bibinfo {author} {\bibfnamefont {F.}~\bibnamefont
  {Joseph~Heremans}}, \bibinfo {author} {\bibfnamefont {L.~C.}\ \bibnamefont
  {Bassett}}, \bibinfo {author} {\bibfnamefont {B.~A.}\ \bibnamefont {Myers}},
  \bibinfo {author} {\bibfnamefont {D.~M.}\ \bibnamefont {Toyli}}, \bibinfo
  {author} {\bibfnamefont {A.~C.}\ \bibnamefont {Bleszynski~Jayich}}, \bibinfo
  {author} {\bibfnamefont {C.~J.}\ \bibnamefont {Palmstrøm}}, \ and\ \bibinfo
  {author} {\bibfnamefont {D.~D.}\ \bibnamefont {Awschalom}},\ }\href {\doibase
  10.1063/1.4748280} {\bibfield  {journal} {\bibinfo  {journal} {Applied
  Physics Letters}\ }\textbf {\bibinfo {volume} {101}},\ \bibinfo {pages}
  {082413} (\bibinfo {year} {2012})}\BibitemShut {NoStop}%
\bibitem [{\citenamefont {Eichhorn}\ \emph {et~al.}(2019)\citenamefont
  {Eichhorn}, \citenamefont {McLellan},\ and\ \citenamefont
  {Bleszynski~Jayich}}]{Eichhorn2019}%
  \BibitemOpen
  \bibfield  {author} {\bibinfo {author} {\bibfnamefont {T.~R.}\ \bibnamefont
  {Eichhorn}}, \bibinfo {author} {\bibfnamefont {C.~A.}\ \bibnamefont
  {McLellan}}, \ and\ \bibinfo {author} {\bibfnamefont {A.~C.}\ \bibnamefont
  {Bleszynski~Jayich}},\ }\href {\doibase 10.1103/PhysRevMaterials.3.113802}
  {\bibfield  {journal} {\bibinfo  {journal} {Phys. Rev. Materials}\ }\textbf
  {\bibinfo {volume} {3}},\ \bibinfo {pages} {113802} (\bibinfo {year}
  {2019})}\BibitemShut {NoStop}%
\bibitem [{\citenamefont {Viola}\ and\ \citenamefont
  {Lloyd}(1998)}]{Viola1998}%
  \BibitemOpen
  \bibfield  {author} {\bibinfo {author} {\bibfnamefont {L.}~\bibnamefont
  {Viola}}\ and\ \bibinfo {author} {\bibfnamefont {S.}~\bibnamefont {Lloyd}},\
  }\href {\doibase 10.1103/PhysRevA.58.2733} {\bibfield  {journal} {\bibinfo
  {journal} {Phys. Rev. A}\ }\textbf {\bibinfo {volume} {58}},\ \bibinfo
  {pages} {2733} (\bibinfo {year} {1998})}\BibitemShut {NoStop}%
\bibitem [{\citenamefont {Viola}\ \emph {et~al.}(1999)\citenamefont {Viola},
  \citenamefont {Knill},\ and\ \citenamefont {Lloyd}}]{Viola1999}%
  \BibitemOpen
  \bibfield  {author} {\bibinfo {author} {\bibfnamefont {L.}~\bibnamefont
  {Viola}}, \bibinfo {author} {\bibfnamefont {E.}~\bibnamefont {Knill}}, \ and\
  \bibinfo {author} {\bibfnamefont {S.}~\bibnamefont {Lloyd}},\ }\href
  {\doibase 10.1103/PhysRevLett.82.2417} {\bibfield  {journal} {\bibinfo
  {journal} {Phys. Rev. Lett.}\ }\textbf {\bibinfo {volume} {82}},\ \bibinfo
  {pages} {2417} (\bibinfo {year} {1999})}\BibitemShut {NoStop}%
\bibitem [{\citenamefont {Bylander}\ \emph {et~al.}(2011)\citenamefont
  {Bylander}, \citenamefont {Gustavsson}, \citenamefont {Yan}, \citenamefont
  {Yoshihara}, \citenamefont {Harrabi}, \citenamefont {Fitch}, \citenamefont
  {Cory}, \citenamefont {Nakamura}, \citenamefont {Tsai},\ and\ \citenamefont
  {Oliver}}]{Bylander2011}%
  \BibitemOpen
  \bibfield  {author} {\bibinfo {author} {\bibfnamefont {J.}~\bibnamefont
  {Bylander}}, \bibinfo {author} {\bibfnamefont {S.}~\bibnamefont
  {Gustavsson}}, \bibinfo {author} {\bibfnamefont {F.}~\bibnamefont {Yan}},
  \bibinfo {author} {\bibfnamefont {F.}~\bibnamefont {Yoshihara}}, \bibinfo
  {author} {\bibfnamefont {K.}~\bibnamefont {Harrabi}}, \bibinfo {author}
  {\bibfnamefont {G.}~\bibnamefont {Fitch}}, \bibinfo {author} {\bibfnamefont
  {D.~G.}\ \bibnamefont {Cory}}, \bibinfo {author} {\bibfnamefont
  {Y.}~\bibnamefont {Nakamura}}, \bibinfo {author} {\bibfnamefont {J.-S.}\
  \bibnamefont {Tsai}}, \ and\ \bibinfo {author} {\bibfnamefont {W.~D.}\
  \bibnamefont {Oliver}},\ }\href {\doibase 10.1038/nphys1994} {\bibfield
  {journal} {\bibinfo  {journal} {Nature Physics}\ }\textbf {\bibinfo {volume}
  {7}},\ \bibinfo {pages} {565} (\bibinfo {year} {2011})}\BibitemShut {NoStop}%
\bibitem [{\citenamefont {de~Lange}\ \emph {et~al.}(2010)\citenamefont
  {de~Lange}, \citenamefont {Wang}, \citenamefont {Rist{\`e}}, \citenamefont
  {Dobrovitski},\ and\ \citenamefont {Hanson}}]{deLange2010}%
  \BibitemOpen
  \bibfield  {author} {\bibinfo {author} {\bibfnamefont {G.}~\bibnamefont
  {de~Lange}}, \bibinfo {author} {\bibfnamefont {Z.~H.}\ \bibnamefont {Wang}},
  \bibinfo {author} {\bibfnamefont {D.}~\bibnamefont {Rist{\`e}}}, \bibinfo
  {author} {\bibfnamefont {V.~V.}\ \bibnamefont {Dobrovitski}}, \ and\ \bibinfo
  {author} {\bibfnamefont {R.}~\bibnamefont {Hanson}},\ }\href {\doibase
  10.1126/science.1192739} {\bibfield  {journal} {\bibinfo  {journal}
  {Science}\ }\textbf {\bibinfo {volume} {330}},\ \bibinfo {pages} {60}
  (\bibinfo {year} {2010})}\BibitemShut {NoStop}%
\bibitem [{\citenamefont {van~der Sar}\ \emph {et~al.}(2012)\citenamefont
  {van~der Sar}, \citenamefont {Wang}, \citenamefont {Blok}, \citenamefont
  {Bernien}, \citenamefont {Taminiau}, \citenamefont {Toyli}, \citenamefont
  {Lidar}, \citenamefont {Awschalom}, \citenamefont {Hanson},\ and\
  \citenamefont {Dobrovitski}}]{Vandersar2012}%
  \BibitemOpen
  \bibfield  {author} {\bibinfo {author} {\bibfnamefont {T.}~\bibnamefont
  {van~der Sar}}, \bibinfo {author} {\bibfnamefont {Z.~H.}\ \bibnamefont
  {Wang}}, \bibinfo {author} {\bibfnamefont {M.~S.}\ \bibnamefont {Blok}},
  \bibinfo {author} {\bibfnamefont {H.}~\bibnamefont {Bernien}}, \bibinfo
  {author} {\bibfnamefont {T.~H.}\ \bibnamefont {Taminiau}}, \bibinfo {author}
  {\bibfnamefont {D.~M.}\ \bibnamefont {Toyli}}, \bibinfo {author}
  {\bibfnamefont {D.~A.}\ \bibnamefont {Lidar}}, \bibinfo {author}
  {\bibfnamefont {D.~D.}\ \bibnamefont {Awschalom}}, \bibinfo {author}
  {\bibfnamefont {R.}~\bibnamefont {Hanson}}, \ and\ \bibinfo {author}
  {\bibfnamefont {V.~V.}\ \bibnamefont {Dobrovitski}},\ }\href {\doibase
  10.1038/nature10900} {\bibfield  {journal} {\bibinfo  {journal} {Nature}\
  }\textbf {\bibinfo {volume} {484}},\ \bibinfo {pages} {82} (\bibinfo {year}
  {2012})}\BibitemShut {NoStop}%
\bibitem [{\citenamefont {Barry}\ \emph {et~al.}(2020)\citenamefont {Barry},
  \citenamefont {Schloss}, \citenamefont {Bauch}, \citenamefont {Turner},
  \citenamefont {Hart}, \citenamefont {Pham},\ and\ \citenamefont
  {Walsworth}}]{Barry2020}%
  \BibitemOpen
  \bibfield  {author} {\bibinfo {author} {\bibfnamefont {J.~F.}\ \bibnamefont
  {Barry}}, \bibinfo {author} {\bibfnamefont {J.~M.}\ \bibnamefont {Schloss}},
  \bibinfo {author} {\bibfnamefont {E.}~\bibnamefont {Bauch}}, \bibinfo
  {author} {\bibfnamefont {M.~J.}\ \bibnamefont {Turner}}, \bibinfo {author}
  {\bibfnamefont {C.~A.}\ \bibnamefont {Hart}}, \bibinfo {author}
  {\bibfnamefont {L.~M.}\ \bibnamefont {Pham}}, \ and\ \bibinfo {author}
  {\bibfnamefont {R.~L.}\ \bibnamefont {Walsworth}},\ }\href {\doibase
  10.1103/RevModPhys.92.015004} {\bibfield  {journal} {\bibinfo  {journal}
  {Rev. Mod. Phys.}\ }\textbf {\bibinfo {volume} {92}},\ \bibinfo {pages}
  {015004} (\bibinfo {year} {2020})}\BibitemShut {NoStop}%
\bibitem [{\citenamefont {Slichter}(1990)}]{Slichter1990}%
  \BibitemOpen
  \bibfield  {author} {\bibinfo {author} {\bibfnamefont {C.~P.}\ \bibnamefont
  {Slichter}},\ }\href {\doibase 10.1007/978-3-662-09441-9} {\emph {\bibinfo
  {title} {Principles of Magnetic Resonance}}},\ \bibinfo {edition} {3rd}\ ed.\
  (\bibinfo  {publisher} {Springer-Verlag Berlin Heidelberg},\ \bibinfo {year}
  {1990})\BibitemShut {NoStop}%
\bibitem [{\citenamefont {Bauch}\ \emph {et~al.}(2018)\citenamefont {Bauch},
  \citenamefont {Hart}, \citenamefont {Schloss}, \citenamefont {Turner},
  \citenamefont {Barry}, \citenamefont {Kehayias}, \citenamefont {Singh},\ and\
  \citenamefont {Walsworth}}]{Bauch2018}%
  \BibitemOpen
  \bibfield  {author} {\bibinfo {author} {\bibfnamefont {E.}~\bibnamefont
  {Bauch}}, \bibinfo {author} {\bibfnamefont {C.~A.}\ \bibnamefont {Hart}},
  \bibinfo {author} {\bibfnamefont {J.~M.}\ \bibnamefont {Schloss}}, \bibinfo
  {author} {\bibfnamefont {M.~J.}\ \bibnamefont {Turner}}, \bibinfo {author}
  {\bibfnamefont {J.~F.}\ \bibnamefont {Barry}}, \bibinfo {author}
  {\bibfnamefont {P.}~\bibnamefont {Kehayias}}, \bibinfo {author}
  {\bibfnamefont {S.}~\bibnamefont {Singh}}, \ and\ \bibinfo {author}
  {\bibfnamefont {R.~L.}\ \bibnamefont {Walsworth}},\ }\href {\doibase
  10.1103/PhysRevX.8.031025} {\bibfield  {journal} {\bibinfo  {journal} {Phys.
  Rev. X}\ }\textbf {\bibinfo {volume} {8}},\ \bibinfo {pages} {031025}
  (\bibinfo {year} {2018})}\BibitemShut {NoStop}%
\bibitem [{\citenamefont {Bluvstein}\ \emph {et~al.}(2019)\citenamefont
  {Bluvstein}, \citenamefont {Zhang}, \citenamefont {McLellan}, \citenamefont
  {Williams},\ and\ \citenamefont {Jayich}}]{Dolev2019}%
  \BibitemOpen
  \bibfield  {author} {\bibinfo {author} {\bibfnamefont {D.}~\bibnamefont
  {Bluvstein}}, \bibinfo {author} {\bibfnamefont {Z.}~\bibnamefont {Zhang}},
  \bibinfo {author} {\bibfnamefont {C.~A.}\ \bibnamefont {McLellan}}, \bibinfo
  {author} {\bibfnamefont {N.~R.}\ \bibnamefont {Williams}}, \ and\ \bibinfo
  {author} {\bibfnamefont {A.~C.~B.}\ \bibnamefont {Jayich}},\ }\href {\doibase
  10.1103/PhysRevLett.123.146804} {\bibfield  {journal} {\bibinfo  {journal}
  {Phys. Rev. Lett.}\ }\textbf {\bibinfo {volume} {123}},\ \bibinfo {pages}
  {146804} (\bibinfo {year} {2019})}\BibitemShut {NoStop}%
\bibitem [{\citenamefont {Ishizu}\ \emph {et~al.}(2020)\citenamefont {Ishizu},
  \citenamefont {Sasaki}, \citenamefont {Misonou}, \citenamefont {Teraji},
  \citenamefont {Itoh},\ and\ \citenamefont {Abe}}]{ishizu2020spin}%
  \BibitemOpen
  \bibfield  {author} {\bibinfo {author} {\bibfnamefont {S.}~\bibnamefont
  {Ishizu}}, \bibinfo {author} {\bibfnamefont {K.}~\bibnamefont {Sasaki}},
  \bibinfo {author} {\bibfnamefont {D.}~\bibnamefont {Misonou}}, \bibinfo
  {author} {\bibfnamefont {T.}~\bibnamefont {Teraji}}, \bibinfo {author}
  {\bibfnamefont {K.~M.}\ \bibnamefont {Itoh}}, \ and\ \bibinfo {author}
  {\bibfnamefont {E.}~\bibnamefont {Abe}},\ }\href@noop {} {\enquote {\bibinfo
  {title} {Spin coherence and depths of single nitrogen-vacancy centers created
  by ion implantation into diamond via screening masks},}\ } (\bibinfo {year}
  {2020}),\ \Eprint {http://arxiv.org/abs/2006.07763} {arXiv:2006.07763
  [quant-ph]} \BibitemShut {NoStop}%
\bibitem [{\citenamefont {Ernst}(1966)}]{Ernst1966}%
  \BibitemOpen
  \bibfield  {author} {\bibinfo {author} {\bibfnamefont {R.~R.}\ \bibnamefont
  {Ernst}},\ }\href {\doibase 10.1063/1.1727409} {\bibfield  {journal}
  {\bibinfo  {journal} {The Journal of Chemical Physics}\ }\textbf {\bibinfo
  {volume} {45}},\ \bibinfo {pages} {3845} (\bibinfo {year}
  {1966})}\BibitemShut {NoStop}%
\bibitem [{sup()}]{supplemental}%
  \BibitemOpen
  \href@noop {} {}\bibinfo {note} {See the Supplemental Material for supporting
  details on sample information, stochastic driving engineering and the spin
  bath model.}\BibitemShut {Stop}%
\bibitem [{\citenamefont {Mamin}\ \emph {et~al.}(2012)\citenamefont {Mamin},
  \citenamefont {Sherwood},\ and\ \citenamefont {Rugar}}]{Mamin2012}%
  \BibitemOpen
  \bibfield  {author} {\bibinfo {author} {\bibfnamefont {H.~J.}\ \bibnamefont
  {Mamin}}, \bibinfo {author} {\bibfnamefont {M.~H.}\ \bibnamefont {Sherwood}},
  \ and\ \bibinfo {author} {\bibfnamefont {D.}~\bibnamefont {Rugar}},\ }\href
  {\doibase 10.1103/PhysRevB.86.195422} {\bibfield  {journal} {\bibinfo
  {journal} {Phys. Rev. B}\ }\textbf {\bibinfo {volume} {86}},\ \bibinfo
  {pages} {195422} (\bibinfo {year} {2012})}\BibitemShut {NoStop}%
\bibitem [{\citenamefont {Grinolds}\ \emph {et~al.}(2014)\citenamefont
  {Grinolds}, \citenamefont {Warner}, \citenamefont {De~Greve}, \citenamefont
  {Dovzhenko}, \citenamefont {Thiel}, \citenamefont {Walsworth}, \citenamefont
  {Hong}, \citenamefont {Maletinsky},\ and\ \citenamefont
  {Yacoby}}]{Grinolds2014}%
  \BibitemOpen
  \bibfield  {author} {\bibinfo {author} {\bibfnamefont {M.~S.}\ \bibnamefont
  {Grinolds}}, \bibinfo {author} {\bibfnamefont {M.}~\bibnamefont {Warner}},
  \bibinfo {author} {\bibfnamefont {K.}~\bibnamefont {De~Greve}}, \bibinfo
  {author} {\bibfnamefont {Y.}~\bibnamefont {Dovzhenko}}, \bibinfo {author}
  {\bibfnamefont {L.}~\bibnamefont {Thiel}}, \bibinfo {author} {\bibfnamefont
  {R.~L.}\ \bibnamefont {Walsworth}}, \bibinfo {author} {\bibfnamefont
  {S.}~\bibnamefont {Hong}}, \bibinfo {author} {\bibfnamefont {P.}~\bibnamefont
  {Maletinsky}}, \ and\ \bibinfo {author} {\bibfnamefont {A.}~\bibnamefont
  {Yacoby}},\ }\href {\doibase 10.1038/nnano.2014.30} {\bibfield  {journal}
  {\bibinfo  {journal} {Nature Nanotechnology}\ }\textbf {\bibinfo {volume}
  {9}},\ \bibinfo {pages} {279–284} (\bibinfo {year} {2014})}\BibitemShut
  {NoStop}%
\bibitem [{\citenamefont {Myers}\ \emph {et~al.}(2014)\citenamefont {Myers},
  \citenamefont {Das}, \citenamefont {Dartiailh}, \citenamefont {Ohno},
  \citenamefont {Awschalom},\ and\ \citenamefont {{Bleszynski
  Jayich}}}]{Myers2014}%
  \BibitemOpen
  \bibfield  {author} {\bibinfo {author} {\bibfnamefont {B.}~\bibnamefont
  {Myers}}, \bibinfo {author} {\bibfnamefont {A.}~\bibnamefont {Das}}, \bibinfo
  {author} {\bibfnamefont {M.}~\bibnamefont {Dartiailh}}, \bibinfo {author}
  {\bibfnamefont {K.}~\bibnamefont {Ohno}}, \bibinfo {author} {\bibfnamefont
  {D.}~\bibnamefont {Awschalom}}, \ and\ \bibinfo {author} {\bibfnamefont
  {A.}~\bibnamefont {{Bleszynski Jayich}}},\ }\href {\doibase
  10.1103/PhysRevLett.113.027602} {\bibfield  {journal} {\bibinfo  {journal}
  {Physical Review Letters}\ }\textbf {\bibinfo {volume} {113}},\ \bibinfo
  {pages} {027602} (\bibinfo {year} {2014})}\BibitemShut {NoStop}%
\bibitem [{\citenamefont {Sangtawesin}\ \emph {et~al.}(2019)\citenamefont
  {Sangtawesin}, \citenamefont {Dwyer}, \citenamefont {Srinivasan},
  \citenamefont {Allred}, \citenamefont {Rodgers}, \citenamefont {De~Greve},
  \citenamefont {Stacey}, \citenamefont {Dontschuk}, \citenamefont {O'Donnell},
  \citenamefont {Hu}, \citenamefont {Evans}, \citenamefont {Jaye},
  \citenamefont {Fischer}, \citenamefont {Markham}, \citenamefont {Twitchen},
  \citenamefont {Park}, \citenamefont {Lukin},\ and\ \citenamefont
  {de~Leon}}]{Sangtawesin2019}%
  \BibitemOpen
  \bibfield  {author} {\bibinfo {author} {\bibfnamefont {S.}~\bibnamefont
  {Sangtawesin}}, \bibinfo {author} {\bibfnamefont {B.~L.}\ \bibnamefont
  {Dwyer}}, \bibinfo {author} {\bibfnamefont {S.}~\bibnamefont {Srinivasan}},
  \bibinfo {author} {\bibfnamefont {J.~J.}\ \bibnamefont {Allred}}, \bibinfo
  {author} {\bibfnamefont {L.~V.~H.}\ \bibnamefont {Rodgers}}, \bibinfo
  {author} {\bibfnamefont {K.}~\bibnamefont {De~Greve}}, \bibinfo {author}
  {\bibfnamefont {A.}~\bibnamefont {Stacey}}, \bibinfo {author} {\bibfnamefont
  {N.}~\bibnamefont {Dontschuk}}, \bibinfo {author} {\bibfnamefont {K.~M.}\
  \bibnamefont {O'Donnell}}, \bibinfo {author} {\bibfnamefont {D.}~\bibnamefont
  {Hu}}, \bibinfo {author} {\bibfnamefont {D.~A.}\ \bibnamefont {Evans}},
  \bibinfo {author} {\bibfnamefont {C.}~\bibnamefont {Jaye}}, \bibinfo {author}
  {\bibfnamefont {D.~A.}\ \bibnamefont {Fischer}}, \bibinfo {author}
  {\bibfnamefont {M.~L.}\ \bibnamefont {Markham}}, \bibinfo {author}
  {\bibfnamefont {D.~J.}\ \bibnamefont {Twitchen}}, \bibinfo {author}
  {\bibfnamefont {H.}~\bibnamefont {Park}}, \bibinfo {author} {\bibfnamefont
  {M.~D.}\ \bibnamefont {Lukin}}, \ and\ \bibinfo {author} {\bibfnamefont
  {N.~P.}\ \bibnamefont {de~Leon}},\ }\href {\doibase
  10.1103/PhysRevX.9.031052} {\bibfield  {journal} {\bibinfo  {journal} {Phys.
  Rev. X}\ }\textbf {\bibinfo {volume} {9}},\ \bibinfo {pages} {031052}
  (\bibinfo {year} {2019})}\BibitemShut {NoStop}%
\bibitem [{\citenamefont {Stacey}\ \emph {et~al.}(2019)\citenamefont {Stacey},
  \citenamefont {Dontschuk}, \citenamefont {Chou}, \citenamefont {Broadway},
  \citenamefont {Schenk}, \citenamefont {Sear}, \citenamefont {Tetienne},
  \citenamefont {Hoffman}, \citenamefont {Prawer}, \citenamefont {Pakes},
  \citenamefont {Tadich}, \citenamefont {de~Leon}, \citenamefont {Gali},\ and\
  \citenamefont {Hollenberg}}]{Stacey2019}%
  \BibitemOpen
  \bibfield  {author} {\bibinfo {author} {\bibfnamefont {A.}~\bibnamefont
  {Stacey}}, \bibinfo {author} {\bibfnamefont {N.}~\bibnamefont {Dontschuk}},
  \bibinfo {author} {\bibfnamefont {J.-P.}\ \bibnamefont {Chou}}, \bibinfo
  {author} {\bibfnamefont {D.~A.}\ \bibnamefont {Broadway}}, \bibinfo {author}
  {\bibfnamefont {A.~K.}\ \bibnamefont {Schenk}}, \bibinfo {author}
  {\bibfnamefont {M.~J.}\ \bibnamefont {Sear}}, \bibinfo {author}
  {\bibfnamefont {J.-P.}\ \bibnamefont {Tetienne}}, \bibinfo {author}
  {\bibfnamefont {A.}~\bibnamefont {Hoffman}}, \bibinfo {author} {\bibfnamefont
  {S.}~\bibnamefont {Prawer}}, \bibinfo {author} {\bibfnamefont {C.~I.}\
  \bibnamefont {Pakes}}, \bibinfo {author} {\bibfnamefont {A.}~\bibnamefont
  {Tadich}}, \bibinfo {author} {\bibfnamefont {N.~P.}\ \bibnamefont {de~Leon}},
  \bibinfo {author} {\bibfnamefont {A.}~\bibnamefont {Gali}}, \ and\ \bibinfo
  {author} {\bibfnamefont {L.~C.~L.}\ \bibnamefont {Hollenberg}},\ }\href
  {\doibase 10.1002/admi.201801449} {\bibfield  {journal} {\bibinfo  {journal}
  {Advanced Materials Interfaces}\ }\textbf {\bibinfo {volume} {6}},\ \bibinfo
  {pages} {1801449} (\bibinfo {year} {2019})}\BibitemShut {NoStop}%
\bibitem [{\citenamefont {de~Sousa}(2009)}]{deSousa2009}%
  \BibitemOpen
  \bibfield  {author} {\bibinfo {author} {\bibfnamefont {R.}~\bibnamefont
  {de~Sousa}},\ }\enquote {\bibinfo {title} {Electron spin as a spectrometer
  of nuclear-spin noise and other fluctuations},}\ in\ \href {\doibase
  10.1007/978-3-540-79365-6_10} {\emph {\bibinfo {booktitle} {Electron Spin
  Resonance and Related Phenomena in Low-Dimensional Structures}}},\ \bibinfo
  {editor} {edited by\ \bibinfo {editor} {\bibfnamefont {M.}~\bibnamefont
  {Fanciulli}}}\ (\bibinfo  {publisher} {Springer Berlin Heidelberg},\ \bibinfo
  {address} {Berlin, Heidelberg},\ \bibinfo {year} {2009})\ pp.\ \bibinfo
  {pages} {183--220}\BibitemShut {NoStop}%
\bibitem [{\citenamefont {Biercuk}\ \emph {et~al.}(2011)\citenamefont
  {Biercuk}, \citenamefont {Doherty},\ and\ \citenamefont {Uys}}]{Biercuk2011}%
  \BibitemOpen
  \bibfield  {author} {\bibinfo {author} {\bibfnamefont {M.~J.}\ \bibnamefont
  {Biercuk}}, \bibinfo {author} {\bibfnamefont {A.~C.}\ \bibnamefont
  {Doherty}}, \ and\ \bibinfo {author} {\bibfnamefont {H.}~\bibnamefont
  {Uys}},\ }\href {\doibase 10.1088/0953-4075/44/15/154002} {\bibfield
  {journal} {\bibinfo  {journal} {Journal of Physics B: Atomic, Molecular and
  Optical Physics}\ }\textbf {\bibinfo {volume} {44}},\ \bibinfo {pages}
  {154002} (\bibinfo {year} {2011})}\BibitemShut {NoStop}%
\bibitem [{\citenamefont {Degen}\ \emph {et~al.}(2017)\citenamefont {Degen},
  \citenamefont {Reinhard},\ and\ \citenamefont {Cappellaro}}]{Degen2017}%
  \BibitemOpen
  \bibfield  {author} {\bibinfo {author} {\bibfnamefont {C.~L.}\ \bibnamefont
  {Degen}}, \bibinfo {author} {\bibfnamefont {F.}~\bibnamefont {Reinhard}}, \
  and\ \bibinfo {author} {\bibfnamefont {P.}~\bibnamefont {Cappellaro}},\
  }\href {\doibase 10.1103/RevModPhys.89.035002} {\bibfield  {journal}
  {\bibinfo  {journal} {Rev. Mod. Phys.}\ }\textbf {\bibinfo {volume} {89}},\
  \bibinfo {pages} {035002} (\bibinfo {year} {2017})}\BibitemShut {NoStop}%
\bibitem [{\citenamefont {Cywi\ifmmode~\acute{n}\else \'{n}\fi{}ski}\ \emph
  {et~al.}(2008)\citenamefont {Cywi\ifmmode~\acute{n}\else \'{n}\fi{}ski},
  \citenamefont {Lutchyn}, \citenamefont {Nave},\ and\ \citenamefont
  {Das~Sarma}}]{Cywinski2008}%
  \BibitemOpen
  \bibfield  {author} {\bibinfo {author} {\bibfnamefont {L.}~\bibnamefont
  {Cywi\ifmmode~\acute{n}\else \'{n}\fi{}ski}}, \bibinfo {author}
  {\bibfnamefont {R.~M.}\ \bibnamefont {Lutchyn}}, \bibinfo {author}
  {\bibfnamefont {C.~P.}\ \bibnamefont {Nave}}, \ and\ \bibinfo {author}
  {\bibfnamefont {S.}~\bibnamefont {Das~Sarma}},\ }\href {\doibase
  10.1103/PhysRevB.77.174509} {\bibfield  {journal} {\bibinfo  {journal} {Phys.
  Rev. B}\ }\textbf {\bibinfo {volume} {77}},\ \bibinfo {pages} {174509}
  (\bibinfo {year} {2008})}\BibitemShut {NoStop}%
\bibitem [{\citenamefont {Tetienne}\ \emph {et~al.}(2018)\citenamefont
  {Tetienne}, \citenamefont {de~Gille}, \citenamefont {Broadway}, \citenamefont
  {Teraji}, \citenamefont {Lillie}, \citenamefont {McCoey}, \citenamefont
  {Dontschuk}, \citenamefont {Hall}, \citenamefont {Stacey}, \citenamefont
  {Simpson},\ and\ \citenamefont {Hollenberg}}]{Tetienne2018}%
  \BibitemOpen
  \bibfield  {author} {\bibinfo {author} {\bibfnamefont {J.-P.}\ \bibnamefont
  {Tetienne}}, \bibinfo {author} {\bibfnamefont {R.~W.}\ \bibnamefont
  {de~Gille}}, \bibinfo {author} {\bibfnamefont {D.~A.}\ \bibnamefont
  {Broadway}}, \bibinfo {author} {\bibfnamefont {T.}~\bibnamefont {Teraji}},
  \bibinfo {author} {\bibfnamefont {S.~E.}\ \bibnamefont {Lillie}}, \bibinfo
  {author} {\bibfnamefont {J.~M.}\ \bibnamefont {McCoey}}, \bibinfo {author}
  {\bibfnamefont {N.}~\bibnamefont {Dontschuk}}, \bibinfo {author}
  {\bibfnamefont {L.~T.}\ \bibnamefont {Hall}}, \bibinfo {author}
  {\bibfnamefont {A.}~\bibnamefont {Stacey}}, \bibinfo {author} {\bibfnamefont
  {D.~A.}\ \bibnamefont {Simpson}}, \ and\ \bibinfo {author} {\bibfnamefont
  {L.~C.~L.}\ \bibnamefont {Hollenberg}},\ }\href {\doibase
  10.1103/PhysRevB.97.085402} {\bibfield  {journal} {\bibinfo  {journal} {Phys.
  Rev. B}\ }\textbf {\bibinfo {volume} {97}},\ \bibinfo {pages} {085402}
  (\bibinfo {year} {2018})}\BibitemShut {NoStop}%
\bibitem [{\citenamefont {Sushkov}\ \emph {et~al.}(2014)\citenamefont
  {Sushkov}, \citenamefont {Lovchinsky}, \citenamefont {Chisholm},
  \citenamefont {Walsworth}, \citenamefont {Park},\ and\ \citenamefont
  {Lukin}}]{Sushkov2014}%
  \BibitemOpen
  \bibfield  {author} {\bibinfo {author} {\bibfnamefont {A.~O.}\ \bibnamefont
  {Sushkov}}, \bibinfo {author} {\bibfnamefont {I.}~\bibnamefont {Lovchinsky}},
  \bibinfo {author} {\bibfnamefont {N.}~\bibnamefont {Chisholm}}, \bibinfo
  {author} {\bibfnamefont {R.~L.}\ \bibnamefont {Walsworth}}, \bibinfo {author}
  {\bibfnamefont {H.}~\bibnamefont {Park}}, \ and\ \bibinfo {author}
  {\bibfnamefont {M.~D.}\ \bibnamefont {Lukin}},\ }\href {\doibase
  10.1103/PhysRevLett.113.197601} {\bibfield  {journal} {\bibinfo  {journal}
  {Phys. Rev. Lett.}\ }\textbf {\bibinfo {volume} {113}},\ \bibinfo {pages}
  {197601} (\bibinfo {year} {2014})}\BibitemShut {NoStop}%
\bibitem [{\citenamefont {Chang}\ \emph {et~al.}(2013)\citenamefont {Chang},
  \citenamefont {Vissers}, \citenamefont {Córcoles}, \citenamefont {Sandberg},
  \citenamefont {Gao}, \citenamefont {Abraham}, \citenamefont {Chow},
  \citenamefont {Gambetta}, \citenamefont {Beth~Rothwell}, \citenamefont
  {Keefe}, \citenamefont {Steffen},\ and\ \citenamefont {Pappas}}]{Chang2013}%
  \BibitemOpen
  \bibfield  {author} {\bibinfo {author} {\bibfnamefont {J.~B.}\ \bibnamefont
  {Chang}}, \bibinfo {author} {\bibfnamefont {M.~R.}\ \bibnamefont {Vissers}},
  \bibinfo {author} {\bibfnamefont {A.~D.}\ \bibnamefont {Córcoles}}, \bibinfo
  {author} {\bibfnamefont {M.}~\bibnamefont {Sandberg}}, \bibinfo {author}
  {\bibfnamefont {J.}~\bibnamefont {Gao}}, \bibinfo {author} {\bibfnamefont
  {D.~W.}\ \bibnamefont {Abraham}}, \bibinfo {author} {\bibfnamefont {J.~M.}\
  \bibnamefont {Chow}}, \bibinfo {author} {\bibfnamefont {J.~M.}\ \bibnamefont
  {Gambetta}}, \bibinfo {author} {\bibfnamefont {M.}~\bibnamefont
  {Beth~Rothwell}}, \bibinfo {author} {\bibfnamefont {G.~A.}\ \bibnamefont
  {Keefe}}, \bibinfo {author} {\bibfnamefont {M.}~\bibnamefont {Steffen}}, \
  and\ \bibinfo {author} {\bibfnamefont {D.~P.}\ \bibnamefont {Pappas}},\
  }\href {\doibase 10.1063/1.4813269} {\bibfield  {journal} {\bibinfo
  {journal} {Applied Physics Letters}\ }\textbf {\bibinfo {volume} {103}},\
  \bibinfo {pages} {012602} (\bibinfo {year} {2013})}\BibitemShut {NoStop}%
\bibitem [{\citenamefont {Brownnutt}\ \emph {et~al.}(2015)\citenamefont
  {Brownnutt}, \citenamefont {Kumph}, \citenamefont {Rabl},\ and\ \citenamefont
  {Blatt}}]{Brownnutt2015}%
  \BibitemOpen
  \bibfield  {author} {\bibinfo {author} {\bibfnamefont {M.}~\bibnamefont
  {Brownnutt}}, \bibinfo {author} {\bibfnamefont {M.}~\bibnamefont {Kumph}},
  \bibinfo {author} {\bibfnamefont {P.}~\bibnamefont {Rabl}}, \ and\ \bibinfo
  {author} {\bibfnamefont {R.}~\bibnamefont {Blatt}},\ }\href {\doibase
  10.1103/RevModPhys.87.1419} {\bibfield  {journal} {\bibinfo  {journal} {Rev.
  Mod. Phys.}\ }\textbf {\bibinfo {volume} {87}},\ \bibinfo {pages} {1419}
  (\bibinfo {year} {2015})}\BibitemShut {NoStop}%
\bibitem [{\citenamefont {Daniilidis}\ \emph {et~al.}(2011)\citenamefont
  {Daniilidis}, \citenamefont {Narayanan}, \citenamefont {Möller},
  \citenamefont {Clark}, \citenamefont {Lee}, \citenamefont {Leek},
  \citenamefont {Wallraff}, \citenamefont {Schulz}, \citenamefont
  {Schmidt-Kaler},\ and\ \citenamefont {Häffner}}]{Daniilidis2011}%
  \BibitemOpen
  \bibfield  {author} {\bibinfo {author} {\bibfnamefont {N.}~\bibnamefont
  {Daniilidis}}, \bibinfo {author} {\bibfnamefont {S.}~\bibnamefont
  {Narayanan}}, \bibinfo {author} {\bibfnamefont {S.~A.}\ \bibnamefont
  {Möller}}, \bibinfo {author} {\bibfnamefont {R.}~\bibnamefont {Clark}},
  \bibinfo {author} {\bibfnamefont {T.~E.}\ \bibnamefont {Lee}}, \bibinfo
  {author} {\bibfnamefont {P.~J.}\ \bibnamefont {Leek}}, \bibinfo {author}
  {\bibfnamefont {A.}~\bibnamefont {Wallraff}}, \bibinfo {author}
  {\bibfnamefont {S.}~\bibnamefont {Schulz}}, \bibinfo {author} {\bibfnamefont
  {F.}~\bibnamefont {Schmidt-Kaler}}, \ and\ \bibinfo {author} {\bibfnamefont
  {H.}~\bibnamefont {Häffner}},\ }\href {\doibase
  10.1088/1367-2630/13/1/013032} {\bibfield  {journal} {\bibinfo  {journal}
  {New Journal of Physics}\ }\textbf {\bibinfo {volume} {13}},\ \bibinfo
  {pages} {013032} (\bibinfo {year} {2011})}\BibitemShut {NoStop}%
\bibitem [{\citenamefont {Hite}\ \emph {et~al.}(2013)\citenamefont {Hite},
  \citenamefont {Colombe}, \citenamefont {Wilson}, \citenamefont {Allcock},
  \citenamefont {Leibfried}, \citenamefont {Wineland},\ and\ \citenamefont
  {Pappas}}]{Hite2013}%
  \BibitemOpen
  \bibfield  {author} {\bibinfo {author} {\bibfnamefont {D.}~\bibnamefont
  {Hite}}, \bibinfo {author} {\bibfnamefont {Y.}~\bibnamefont {Colombe}},
  \bibinfo {author} {\bibfnamefont {A.}~\bibnamefont {Wilson}}, \bibinfo
  {author} {\bibfnamefont {D.}~\bibnamefont {Allcock}}, \bibinfo {author}
  {\bibfnamefont {D.}~\bibnamefont {Leibfried}}, \bibinfo {author}
  {\bibfnamefont {D.}~\bibnamefont {Wineland}}, \ and\ \bibinfo {author}
  {\bibfnamefont {D.}~\bibnamefont {Pappas}},\ }\href {\doibase
  10.1557/mrs.2013.207} {\bibfield  {journal} {\bibinfo  {journal} {MRS
  Bulletin}\ }\textbf {\bibinfo {volume} {38}},\ \bibinfo {pages} {826–833}
  (\bibinfo {year} {2013})}\BibitemShut {NoStop}%
\bibitem [{\citenamefont {Noel}\ \emph {et~al.}(2019)\citenamefont {Noel},
  \citenamefont {Berlin-Udi}, \citenamefont {Matthiesen}, \citenamefont {Yu},
  \citenamefont {Zhou}, \citenamefont {Lordi},\ and\ \citenamefont
  {H\"affner}}]{Noel2019}%
  \BibitemOpen
  \bibfield  {author} {\bibinfo {author} {\bibfnamefont {C.}~\bibnamefont
  {Noel}}, \bibinfo {author} {\bibfnamefont {M.}~\bibnamefont {Berlin-Udi}},
  \bibinfo {author} {\bibfnamefont {C.}~\bibnamefont {Matthiesen}}, \bibinfo
  {author} {\bibfnamefont {J.}~\bibnamefont {Yu}}, \bibinfo {author}
  {\bibfnamefont {Y.}~\bibnamefont {Zhou}}, \bibinfo {author} {\bibfnamefont
  {V.}~\bibnamefont {Lordi}}, \ and\ \bibinfo {author} {\bibfnamefont
  {H.}~\bibnamefont {H\"affner}},\ }\href {\doibase 10.1103/PhysRevA.99.063427}
  {\bibfield  {journal} {\bibinfo  {journal} {Phys. Rev. A}\ }\textbf {\bibinfo
  {volume} {99}},\ \bibinfo {pages} {063427} (\bibinfo {year}
  {2019})}\BibitemShut {NoStop}%
\bibitem [{\citenamefont {Myers}\ \emph {et~al.}(2017)\citenamefont {Myers},
  \citenamefont {Ariyaratne},\ and\ \citenamefont {Jayich}}]{Myers2017}%
  \BibitemOpen
  \bibfield  {author} {\bibinfo {author} {\bibfnamefont {B.~A.}\ \bibnamefont
  {Myers}}, \bibinfo {author} {\bibfnamefont {A.}~\bibnamefont {Ariyaratne}}, \
  and\ \bibinfo {author} {\bibfnamefont {A.~C.~B.}\ \bibnamefont {Jayich}},\
  }\href {\doibase 10.1103/PhysRevLett.118.197201} {\bibfield  {journal}
  {\bibinfo  {journal} {Phys. Rev. Lett.}\ }\textbf {\bibinfo {volume} {118}},\
  \bibinfo {pages} {197201} (\bibinfo {year} {2017})}\BibitemShut {NoStop}%
\end{thebibliography}%


%
\end{document}